\documentclass[sigplan,screen]{acmart}

\usepackage{stfloats}
\usepackage{lipsum}
\usepackage{xtab}
\AtBeginDocument{%
  \providecommand\BibTeX{{%
    \normalfont B\kern-0.5em{\scshape i\kern-0.25em b}\kern-0.8em\TeX}}}

\usepackage{ltablex,booktabs}

\acmConference[]{The Local Meeting for Open Challenges in Autonomous Systems}{November 2023}{Tehran, Iran}

\begin{document}
\title{An Introduction to Adaptive Software Security}

\author{Mehran Alidoost Nia}
\email{alidoostnia@ut.ac.ir}
\orcid{0000-0002-7274-9569}
\affiliation{%
  \institution{School of Electrical and Computer Engineering, University of Tehran}
  \city{Tehran}
  \state{Tehran}
  \country{Iran}
}

\renewcommand{\shortauthors}{Mehran Alidoost Nia}

\begin{abstract}
This paper presents the adaptive software security model, an innovative approach integrating the MAPE-K loop and the Software Development Life Cycle (SDLC). It proactively embeds security policies throughout development, reducing vulnerabilities from different levels of software engineering. Three primary contributions—MAPE-K integration, SDLC embedding, and analytical insights—converge to create a comprehensive approach for strengthening software systems against security threats. This research represents a paradigm shift, adapting security measures with agile software development and ensuring continuous improvement in the face of evolving threats. The model emerges as a robust solution, addressing the crucial need for adaptive software security strategies in modern software development. We analytically discuss the advantages of the proposed model. 
\end{abstract}


\maketitle

\section{Introduction}
Functional and non-functional requirements are both crucial to the successful operation of software systems. Functional requirements refer to the specific features that a software system is intended to perform, while non-functional requirements relate to how the system's functionalities are evaluated. While it is relatively straightforward to ensure that these requirements are met during the design phase using formal methods or rigorous static analysis techniques, ensuring compliance at runtime is more complex, as changes are expected. To meet these requirements at runtime, it is necessary to capture changes and respond appropriately to each shift. This adaptivity is essential to ensure that the system's requirements are met and that it continues to function optimally according to those requirements.

\begin{figure*}[h]

  \centering
  \includegraphics[width=\textwidth]{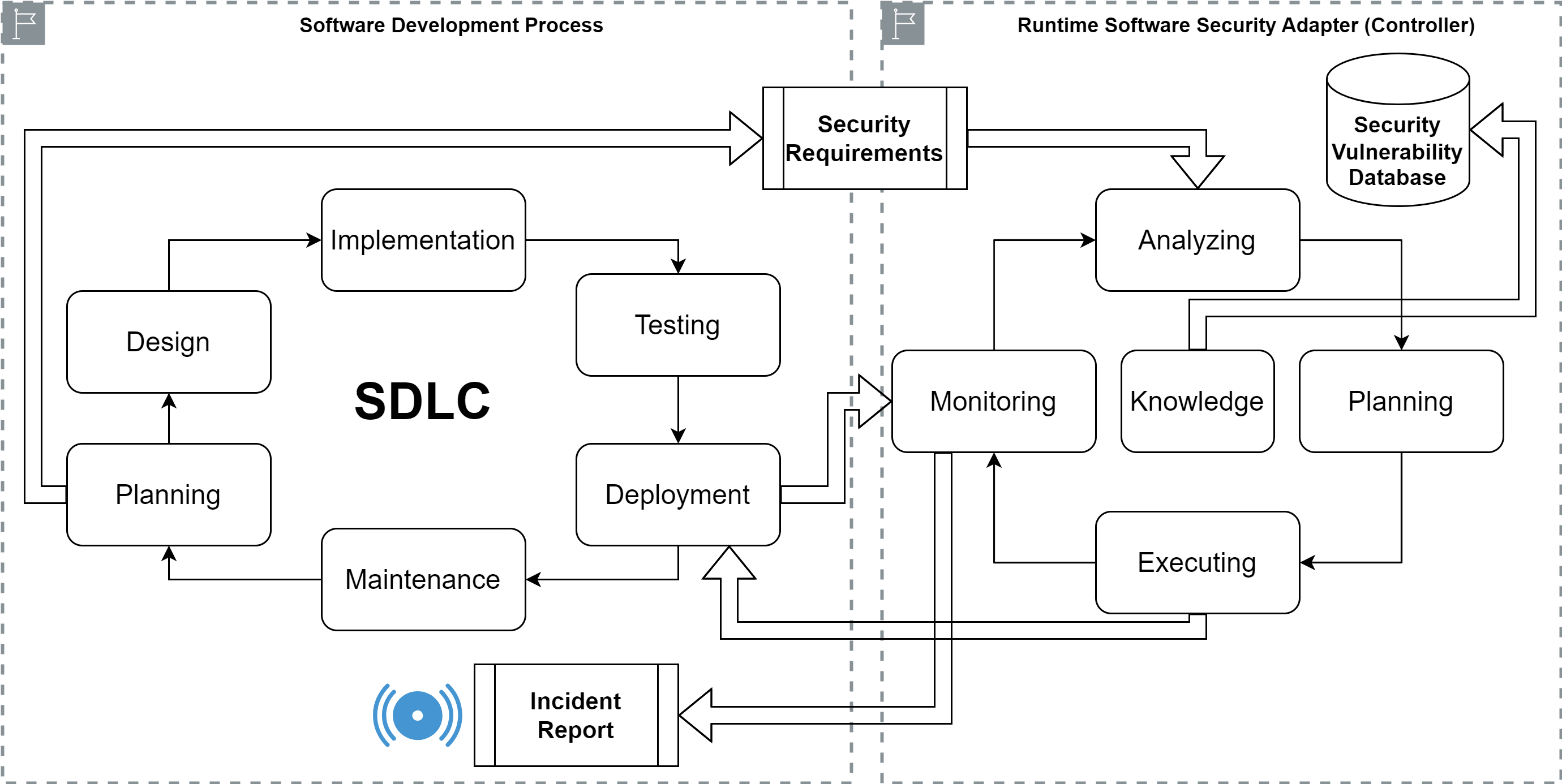}
  \caption{Overview of the proposed adaptive software security model: It composes from SDLC process, the MAPE-K loop and security components.}
  \label{fig:sdlc}
\end{figure*}

In software development, it is important to consider non-functional requirements such as reliability, usability, security, safety, and performance. These requirements do not relate to the functionality of the system but rather to its non-functional aspects. The software development life-cycle (SDLC) provides a framework for the development and deployment of software~\cite{sdlc1}, as shown in Figure~\ref{fig:sdlc}. Security requirements are crucial to ensure the production of secure software. Software security requirements include availability, confidentiality, and integrity of the software system. These requirements specifically address potential security risks and vulnerabilities and are important to consider in the overall software requirements. Depending on the situations and application requirements, these security requirements may be subject to change at runtime.

When it comes to adaptive software development, the software must have runtime strategies to handle situations where a few configurations or user-related assumptions change. For example, the communication protocol may change suddenly, and the defense strategies should be adapted to the system's new situation. Under all circumstances, the security requirements of the system must be fulfilled~\cite{sdlc-sec1}. This is the reason behind any adaptation tactics we perform at runtime. The goal of incorporating security requirements into the software development process is to design, implement, and maintain a system that can resist and mitigate various security threats. The SDLC process must be changed to cover security-related requirements at runtime~\cite{sdlc2}. In the software development process, by the runtime, we mean the environment which the software works within that after deployment. 

Consider a web server that tolerates many sensitive requests from users. The server's security must be guaranteed even if parts of the requests are malicious. When the server is under attack, the system must adaptively detect the new situation and plan for possible changes in its strategies at runtime. The goal is to preserve all three security requirements of the users and their data. In such situations, the system may restrict the users' IP addresses to where malicious requests belong. Another adaptation security action could be restricting the number of requests issued from a unique user. All these adaptation actions are performed to ensure security requirements at runtime when changes come to violate those requirements. 

We intend to extend the adaptive software security underlying to self-adaptive software systems where reaction(s) to changes at runtime are performed autonomously by a well-known mechanism called MAPE loop~\cite{mape}. MAPE stands for Monitoring, Analysing, Planning, and Executing the adaptation actions at runtime of a software system. Learning from the environment and possible adaptation actions moves the self-adaptive software systems toward using a knowledge-based MAPE loop model called MAPE-K~\cite{man1}. In this model, the software system learns from changes affecting security requirements and tries to use previous experience to alleviate the effect of changes in the software security strategies. A MAPE-based self-adaptive system can support the system's security requirements to be better monitored at runtime and capture the changes by analytical and rigorous techniques. Then, the software system can plan and execute the required adaptation actions periodically. 

In this paper, we outline the essence of adaptive software security, categorize the applications, analyze the intersection between the SDLC and MAPE loop in the context of software security requirements~\cite{sdlc-sec2}, and give a few analytical insights into research opportunities. 

The rest of the paper is as follows. In the next section, we discuss the required background and underlying sub-systems. Section Three discusses the model for adaptive software security and the security requirements at runtime. Section Four is dedicated to incorporating the SDLC model and MAPE loop in the context of adaptive software security. Section Five analyzes the applications and advantages of adaptive software security in self-adaptive software systems' safe and secure autonomy. Finally, in Section Six, we conclude the paper.

\section{Literature Review}
In this section, we discuss the required background, literature and technical aspects of adaptive software design, requirement engineering of adaptive software systems and adaptive secure software systems.

\subsection{Software Design Life Cycle (SDLC)}
The Software Development Life Cycle (SDLC) is a systematic and structured approach to developing, deploying, and maintaining software applications. It encompasses a series of phases that guide the entire software development process. Understanding the fundamentals of SDLC is crucial for establishing a solid foundation in software engineering and ensuring the delivery of high-quality, secure, and reliable software systems. The SDLC typically consists of several key phases, each serving a specific purpose in the development process. The first phase is the Planning Phase, where project objectives, requirements, and constraints are defined. This phase lays the groundwork for subsequent activities by establishing a clear roadmap for development. Following planning is the Analysis Phase, during which detailed requirements (both functional and non-functional) are gathered, analyzed, and documented~\cite{sdlc4}. 

The Design Phase translates the gathered requirements into an architectural blueprint for the software. It involves creating architectural and detailed designs, specifying how the software will be structured, and defining the relationships between its components. This phase is crucial for establishing the groundwork for both functionality and security.

Once the design is complete, the development team moves to the Implementation Phase, where actual coding occurs~\cite{sdlc5}. Developers write code based on the design specifications, and the software begins to take shape. This phase is about coding and adhering to coding standards, implementing security best practices, and conducting regular code reviews to ensure quality and security. After implementation is the Testing Phase, where the software undergoes rigorous testing to identify defects and vulnerabilities. 

After successful testing, the software advances to the Deployment Phase, where it is released to users. This phase involves careful planning to minimize disruptions and ensure a smooth transition from development to production. Security considerations during deployment include configuring secure settings, validating access controls, and implementing encryption where necessary. The final phase is Maintenance, where the software is monitored, updated, and improved over its operational life cycle. This phase addresses bug fixes and security patches, incorporating new features or changes based on user feedback and evolving requirements. By adhering to the fundamentals of the Software Development Life Cycle, organizations can systematically develop software solutions that meet software requirements and adhere to security best practices throughout the entire development process. This structured approach enhances software systems' reliability, maintainability, and security.

SDLC and software security share a close relationship, where each phase of the SDLC offers opportunities to integrate and enhance security measures. Security considerations are systematically integrated from the early stages of planning and analysis, where security requirements are identified, to the later phases of design, implementation, testing, deployment, and maintenance. The SDLC serves as a framework for weaving security into the fabric of software development, emphasizing the importance of addressing vulnerabilities at every stage. Organizations can proactively strengthen their software systems by adopting a comprehensive approach that aligns security practices with the entire SDLC, ensuring resilience in the face of emerging threats throughout the software's life cycle.

\subsection{Self-Adaptive Software Systems}
Self-adaptive software systems revolve around creating intelligent, responsive software that can autonomously adjust its behavior in response to changing conditions, requirements, and threats, often at runtime. The MAPE-K loop is at the core of these systems, a conceptual framework that guides the self-adaptation process~\cite{mant}. The MAPE-K loop consists of four main components: Monitoring, Analysis, Planning, and Execution, with an additional Knowledge component.

The Monitoring phase involves continuously observing the software system and its environment to collect relevant data. By environment, we mean everything except the controller or the controlled system that can effect the functionalities of the system, e.g., user load or temperature~\cite{pla-sdp}. The monitoring data may include performance metrics, environmental variables, and information about potential security threats. Monitoring serves as the eyes and ears of the adaptive system, providing real-time insights into the software's operational context.

In the Analyzing phase, the collected data is processed to discern patterns~\cite{man3}, identify anomalies, and assess the system's current state. This phase involves leveraging various analytical techniques, including model-based techniques and machine learning algorithms~\cite{ML-sdlc2}, to understand the implications of observed changes and potential security risks. The goal is to understand the system's behavior and security posture comprehensively.

The Planning phase utilizes the insights gained from analysis to formulate adaptive strategies~\cite{model}. This involves determining how the software should adapt to mitigate vulnerabilities, enhance performance, or respond to changing requirements. The planning phase aims to generate actionable plans of the software system, considering both functional and security-related objectives.

In the Executing phase, the software system implements the planned adaptations. This may involve reconfiguring system parameters, adjusting security settings, or deploying updates to address identified vulnerabilities. The execution phase ensures the scheduled adaptations are seamlessly integrated into the operational environment without disrupting the system's functionality.

The Knowledge component represents the intelligence of the adaptive system~\cite{simos3}. It encapsulates the accumulated knowledge about the system, its environment, and past adaptations experiences. This knowledge is crucial for informing future adaptive decisions and refining adaptive strategies. By learning from past experiences, the adaptive system becomes more skillful at making informed decisions that align with the evolving needs and challenges of the software environment.

The self-adaptive software systems center on the MAPE-K loop, a cyclical process that enables software to Monitor its environment, Analyze collected data, Plan adaptations, Execute changes, and continually accumulate knowledge for future decision-making. This framework empowers software systems to autonomously and intelligently respond to dynamic conditions, making them inherently more resilient and secure in the face of ever-evolving challenges. There are many implementations of self-adaptive systems and MAPE loop like~\cite{rep1,rep2}

\subsection{Adaptive Software Design and Security}

Adaptive software design, implementation, and deployment represent a transformative paradigm in software engineering, aiming to enhance the resilience and security of software systems in the face of evolving threats and dynamic environments. In response to the ever-changing landscape of cyber threats, traditional static security measures have proven insufficient, necessitating adopting adaptive approaches to fortify software against emerging vulnerabilities~\cite{runsec}. In the domain of adaptive software design, the emphasis is on creating systems that can autonomously adjust their behavior and configurations in real-time based on the analysis of contextual information and threat intelligence. This approach involves the integration of adaptive algorithms and machine learning techniques~\cite{ML-sdlc}, enabling software to learn from its operational environment, identify potential risks, and proactively adapt to mitigate vulnerabilities. The goal is to create software that resists known attacks and possesses the agility to respond effectively to novel and sophisticated threats.

The implementation of adaptive software involves incorporating dynamic security controls, self-monitoring mechanisms, and modular architectures that facilitate runtime updates. Adaptive software can continuously assess its security posture and adjust its defenses by leveraging techniques such as runtime code analysis, anomaly detection, and behavioral profiling. This dynamic adaptability is crucial for addressing vulnerabilities during runtime, ensuring that the software remains robust against unforeseen exploits.

The deployment phase of adaptive software involves considerations for the seamless integration of security updates and patches. Continuous deployment pipelines, containerization, and microservices architectures are pivotal in rapidly disseminating security enhancements. Additionally, adaptive deployment strategies enable organizations to implement phased releases, allowing for careful monitoring of software behavior in diverse environments before widespread deployment. This iterative approach helps identify potential issues early on and allows for rapid responses to emerging threats.

\section{Adaptive Software Security Model}
In this section, we review adaptive software security strategies against possible environmental changes, and dive into the MAPE loop to incorporate it with software security paradigms. 

\subsection{Security Threats and Adaptation Actions}

Adaptive software security refers to the ability of a software system to dynamically adjust and respond to changing security threats and challenges. It involves the implementation of mechanisms and strategies that allow the software to adapt its security measures at runtime based on the evolving threat landscape coming from the environment~\cite{sdlc-sec3}.

Assume one is developing an online shop platform where users can make online purchases. The system initially employs traditional security measures such as firewalls, encryption, and access controls to protect user data and financial transactions. However, the threat landscape is dynamic, with new vulnerabilities and attack vectors emerging regularly. For example, detecting new generation of threats matters~\cite{newgen} when unknown cyber threats evolves and our security database has no certain knowledge about them. 

In an adaptive software security framework, the software system continuously monitors its environment for potential security threats and adapts its security posture accordingly~\cite{secself}. We review how this could work in the context of the running example. Consider that an adaptive software system follows the structure of a MAPE-K loop introduced in Section 2. The most important parts of the MAPE-K loop are monitoring and analyzing steps where security threats are detected and the situation is analyzed for decision-making. If the system decides to change its current policies, the security strategy of the system is upgraded at runtime according to the severity of threats. Types of adaptive software security actions are categorized as follows:

\begin{itemize}
    \item Behavioral Analysis: The system incorporates behavioral analysis tools to monitor user activities and detect anomalous behavior~\cite{analytical}. For instance, if a user suddenly starts making multiple high-value transactions in a short period, the system may adapt by triggering additional authentication steps.
    \item Threat Intelligence Integration: The platform integrates with external threat intelligence feeds to stay updated on the latest security threats. If a new type of attack becomes prevalent~\cite{privacy, newgen}, the system adapts its intrusion detection and prevention mechanisms to counteract the specific threat, e.g., blocking suspicious ranges of IP addresses.
    \item Dynamic Access Controls: The system employs dynamic access controls that adjust user permissions based on contextual information. For example, if a user typically accesses the platform from a specific geographical location and suddenly attempts to log in from a different country, the system may adapt by enforcing additional verification steps, e.g., triggering a captcha mechanism to assure the user is not a robot.
    \item Automated Patching and Updates: To address known vulnerabilities, the platform implements automated patching and updates. When a critical security patch is released, the system adapts by applying the patch promptly to eliminate potential exploits. For example, the system applies zero day vulnerabilities patches as soon as possible. 
    \item Machine Learning for Anomaly Detection: The platform utilizes machine learning algorithms to analyze patterns in user behavior and network traffic. If the system identifies patterns indicative of a potential security threat, it adapts by adjusting its security rules or triggering alerts for further investigation. In such cases, real-time adaptive actions may not be possible. Decisions must be made in analyzing step of the MAPE-K loop, and the knowledge base should be updated accordingly. In the next steps, it is possible to use this knowledge for planning appropriate adaptation actions.
    \item User Education and Phishing Protection: The system integrates features to educate users about phishing risks. If a new phishing technique becomes prevalent, the platform adapts by enhancing its user education materials and implementing additional safeguards against phishing attempts. As new threats become prevalent, the educational materials are updated, respectively.
\end{itemize}

\subsection{Adaptive Software Security through MAPE Loop}

After introducing possible adaptation actions against environmental changes (security threats), we need to incorporate the software security paradigms with the standard MAPE loop structure.

The MAPE loop is a control loop widely used in autonomic computing systems for managing and adapting to changes in the environment. We discuss each element of the model in the context of software security, focusing on the runtime aspects:

A) Monitoring
\begin{itemize}
    \item Security Event Monitoring: We continuously monitor the software system for security events, including user activities, system logs, network traffic, and any other relevant security-related data.
    \item Behavioral Analysis: We should employ tools and algorithms for real-time behavioral analysis to identify anomalies and potential security threats.
    \item Threat Intelligence Integration: We integrate the software system with external threat intelligence feeds to gather information about the latest security threats and vulnerabilities.
\end{itemize}
B) Analyzing:
\begin{itemize}
    \item Security of Information and Event Management (SIEM): In this step, on can aggregate and correlate security events to derive meaningful insights. We use SIEM tools to analyze patterns, detect anomalies, and identify potential security incidents.
    \item Machine Learning Algorithms: Using the AI power, the system can apply machine learning algorithms for pattern recognition and anomaly detection. We train models on historical data to improve the accuracy of identifying new and evolving threats.
    \item Knowledge Extraction: The MAPE loop extracts knowledge from security incidents and responses to contribute to the system's knowledge base. This includes understanding the effectiveness of previously issued responses and adapting the model based on lessons learned. Note that changing model at runtime is not an easy work~\cite{man4}.
\end{itemize}
C) Planning
\begin{itemize}
    \item Risk Assessment: The system evaluates the severity and potential impact of identified security threats, and assigns risk scores to prioritize responses based on the criticality of the threats.
    \item Dynamic Access Controls: We develop plans for adjusting access controls dynamically based on the detected threats and contextual information.
    \item Adaptive Policies: We may define adaptive security policies that can be activated in response to a set of specific security events. In this step, it is recommended to incorporate knowledge gained from past incidents to refine and improve response strategies.
\end{itemize}
D) Executing
\begin{itemize}
    \item Automated Response: To implement automated responses for mitigating the identified security threats. For example, on may deploy additional authentication measures, update firewall rules, or isolate compromised components.
    \item Dynamic Patching and Updates: The system executes plans for applying security patches and updates at runtime to address known vulnerabilities.
    \item User Notifications: The system needs to communicate security-related information to users, providing guidance and alerts when necessary, and we use knowledge from past incidents to enhance the clarity and effectiveness of communication.
\end{itemize}
E) Knowledge (Learning)
\begin{itemize}
    \item Continuous Learning: The adaptive software system should capture knowledge from each iteration of the loop, including the effectiveness of security responses, the accuracy of threat predictions, and the outcomes of security actions.
    \item Adaptation and Improvement: The system uses the accumulated knowledge to adapt and improve the security model over time. This may involve refining machine learning models, updating threat intelligence, and enhancing response strategies based on historical data.
\end{itemize}

At runtime, the MAPE-K loop emphasizes not only the immediate response to security events but also the continuous learning and adaptation of the software security system based on the knowledge gained from past experiences. This iterative process contributes to a more intelligent and resilient adaptive software security model. However, the success of an adaptive security model depends on the accuracy of monitoring, the effectiveness of analysis and decision-making, and the efficiency of response mechanisms, all of which contribute to a more resilient and adaptive software security system.

\section{Incorporating SDLC with Adaptive Software Security Model}

 Combining the Software Development Life Cycle (SDLC) with the Adaptive Software Security Model (presented in Section 3) and the MAPE-K loop creates a comprehensive approach to building, maintaining, and continuously improving secure software at runtime. In this section, we define the incorporated model and describe each elements. First, we start by diving into the significant role of the SDLC. 

 \subsection{Secure Adaptive SDLC Model}
 In this part, we adapt and redefine SDLC from adaptive software security model point of view. 
 \vspace{3px}

\noindent A) Initiation: Defining Security Goals
\vspace{3px}

In this step of SDLC, we should identify and define security goals during the project initiation phase. The nature of the application, potential threats, and regulatory requirements must be considered in this stage.
\vspace{3px}

\noindent B) Planning: Security Requirements and Design
\vspace{3px}

In this step, We define security requirements based on the identified goals. To better use the requirements, we incorporate adaptive security measures into the design phase, considering the dynamic nature of threats. As discussed earlier in Section 3, the security requirements must be positioned in MAPE-K loop steps.
\vspace{3px}

\noindent C) Development: Secure Coding and Testing
\vspace{3px}

In SDLC, implementation of best development practices is significant. In the development stage, we use security best practice, implement secure coding and conduct security testing during development. Developers must integrate adaptive security mechanisms within the codebase to allow the software system for real-time adjustments against security incidents at runtime.
\vspace{3px}

\noindent D) Testing: Security Testing and Evaluation
\vspace{3px}

In this step of SDLC, several security testing techniques can be applied. One can conduct thorough security testing, including static analysis, dynamic analysis, and penetration testing. SDLC uses the MAPE-K loop during testing to refine security measures based on observed behaviors and potential vulnerabilities.
\vspace{3px}

\noindent E) Deployment: Continuous Monitoring Setup
\vspace{3px}

Continuous delivery is an important notion in SDLC. To best fit this step with the adaptive software security model, we can implement continuous monitoring tools and mechanisms during deployment. SDLC activates the MAPE-K loop to monitor the dynamic system, capturing real-time security events and behaviors.
\vspace{3px}

\noindent F) Operation and Maintenance: Runtime Adaptive Security
\vspace{3px}

To use the runtime benefits in SDLC model, we can activate the MAPE-K loop during the operational phase. The system can continuously monitor, analyze, plan, execute, and learn from security events, adapting security measures in response to evolving threats at runtime.
\vspace{3px}

\noindent G) Knowledge Transfer: Federated Knowledge
\vspace{3px}

The entire SDLC process must capture knowledge gained from the MAPE-K loop and incorporate it into the organization's knowledge base for software systems. The adaptive software security model then uses this knowledge to enhance security practices in future projects.

\subsection{Secure MAPE-K Loop Integration with Security Paradigms}

\vspace{3px}

\noindent A) Security Monitoring
\vspace{3px}

The adaptive software system must use continuous monitoring of security events using adaptive mechanisms to capture any security-related changes. Extracting knowledge from monitored events and behaviors is necessary.
\vspace{3px}

\noindent B) Security Analyzing
\vspace{3px}

To ensure security requirements at runtime, the adaptive software system applies machine learning and analysis to identify threat patterns and anomalies. Incorporating knowledge extraction from historical security incidents is also performed at this stage.
\vspace{3px}

\noindent C) Security Planning
\vspace{3px}

The adaptive software system creates adaptive security plans based on risk assessments. In this step, the adaptive system utilizes knowledge gained from past incidents to enhance response plans.
\vspace{3px}

\noindent D) Security Execution
\vspace{3px}

After precise analysis and planning the security adaptation actions, the adaptive software system must include automated responses and adjustments to security configurations. Executing plans for dynamic access controls and updates based on real-time analysis is done at this step of the MAPE-K loop.
\vspace{3px}

\noindent E) Security Knowledge
\vspace{3px}

One of the main duties of an adaptive software security system is to continuously learn and adapt based on knowledge gained.
Refining the adaptive security model using historical data is of significance.

\section{Analytical Discussions}

This cohesive adaptive software security model establishes a synergistic approach, seamlessly integrating security as an inherent element within the software development process. By embedding security considerations throughout the entire SDLC~\cite{sdlc8}, the model ensures that proactive measures are taken at every stage, from initial planning to deployment and ongoing maintenance. The adaptive nature of the model contributes to the system's resilience, allowing it to dynamically respond and adapt to emerging security threats~\cite{sec-sdlc4}. This adaptive stance goes beyond conventional security measures, fostering a robust defense against the constantly evolving landscape of potential risks. In essence, the model positions security not merely as an added layer but as an integral and adaptive component, fortifying the system against the challenges posed by ever-changing security landscapes. In this section, we comprehensively discuss each of the benefits associated with the integrated model that combines the Software Development Life Cycle (SDLC)~\cite{sdlc10}, the Adaptive Software Security Model, and the MAPE-K loop:

A) Proactive Security
\begin{itemize}
    \item Integrated Security Throughout SDLC: Instead of considering security as a separate phase, the integrated model ensures that security requirements are embedded in every stage of the SDLC~\cite{sec-sdlc5}. This includes planning, design, development, testing, deployment, and maintenance.
    \item Risk Mitigation from the Start: Proactive security means identifying and addressing potential security risks early in the development process. This approach helps prevent vulnerabilities from being introduced and minimizes the likelihood of security issues in the final product in its runtime.
    \item Cost-Effective Security Measures: Addressing security concerns during the early stages of development is often more cost-effective than attempting to incorporate security into a system that has already been designed and implemented without proper security considerations.
    \item Alignment with Best Practices: The model encourages adherence to security best practices and standards throughout the development life cycle, utilizing a security-aware culture within the development team.
\end{itemize}

B) Continuous Improvement
\begin{itemize}
    \item MAPE-K Loop for Learning and Adaptation: The MAPE-K loop continuously monitors, analyzes, plans, executes, and learns from security incidents. This iterative process facilitates continuous improvement by incorporating insights from previous security events.
    \item Adaptive Security Strategies: Continuous improvement involves adapting security strategies based on evolving threats and changing circumstances. The model enables the organization to learn from both successful security measures and incidents, refining its approach over time as the MAPE-K loop executes.
    \item Optimization of Security Measures: The continuous learning aspect of the model allows for the improvements of security measures. As the system collects knowledge about threat patterns, vulnerabilities, and effective responses, it can fit security protocols for vast efficiency.
    \item Agile Response to Emerging Threats: In a rapidly evolving software security landscape, continuous improvement ensures that the organization remains agile in responding to new and emerging threats recognized in software systems. Security measures are not static but evolve based on the latest intelligence and experiences gained by the iterative process of the MAPE-K loop.
\end{itemize}

C) Reduced Time-to-Response:
\begin{itemize}
    \item Runtime Adaptive Security Measures: The model incorporates runtime adaptive security measures through the MAPE-K loop, enabling quick responses to security incidents and emerging threats.
    \item Minimization of Impact: Rapid responses to security incidents help minimize the impact of potential security breaches. By identifying and mitigating threats at runtime, the software system can prevent or limit damage to systems, data, and user trust.
    \item Enhanced Incident Response: Reduced time-to-response improves the effectiveness of incident response efforts. The organization can isolate compromised software components, apply security patches promptly, and implement corrective actions rapidly.
    \item Improved Resilience: Runtime adaptive security measures enhance the system's resilience by addressing vulnerabilities and threats once they are monitored. This reduces the attack time for intruders and strengthens the overall software security fundamentals.
\end{itemize}

D) Informed Decision-Making:
\begin{itemize}
    \item Knowledge Extraction and Transfer: The adaptive software security model emphasizes capturing and transferring knowledge gained from security events, responses, and historical data.
    \item Evidence-Based Decision-Making: Knowledge extracted from precedents incidents serves as a valuable knowledge base for decision-making. It gives insights into the effectiveness of security measures, allowing the team to make informed choices.
    \item Risk-Based Decision-Making: The organization can assess risks more accurately by leveraging historical data and security logs. This enables the software security team to prioritize security measures based on the similarities~\cite{similarity} and potential impact of different threats.
    \item Strategic Planning: Informed decision-making extends beyond immediate responses to security incidents. The organization can use accumulated knowledge to inform long-term strategic planning, such as resource allocation, security architecture improvements, and integrated security policy development.
\end{itemize}

As shown in Table~\ref{tab:metrics}, we give a list of evaluation metrics for validating the proposed adaptive software security model after implementation. The metrics in both SDLC and MAPE-K loop important for measuring the outcome of the software engineering models~\cite{sdlc6}. The monitoring phase involves repeatedly collecting relevant data for each metric. For the rate of anomalies, it monitors runtime anomaly occurrences. For time-to-response, it tracks the time taken to address security threats. Dynamic access controls observe user behavior and contextual changes. 

During the analyzing phase, the system analyzes the monitored data to identify patterns, trends, and potential security threats. For each metric, this stage aims to understand the current state and detect anomalies, urgency, or effectiveness of existing policies. The analysis phase informs subsequent decision-making. The planning phase develops strategies for adapting security measures based on the analysis. The rate of anomalies may involve adjusting anomaly detection thresholds. For time-to-response, it formulates plans to expedite incident response. For dynamic access controls, it plans adjustments based on contextual changes. 

In the executing phase, the system carries out the planned strategies at runtime. The execution includes deploying adaptive responses, adjusting access controls, or responding to incidents promptly. The effectiveness of the execution is crucial in addressing and mitigating security threats. Consequently, the learning phase captures insights and outcomes from executing adaptive strategies. It involves learning from the success or failure of responses. For each metric, this stage contributes to continuously improving the system's adaptive security capabilities and refining strategies for the future.

\section{Conclusion and the Future Direction}

In conclusion, the Adaptive Software Security Model presents three pivotal contributions to cybersecurity. Firstly, seamless integration of the MAPE-K loop enhances runtime and autonomous adaptability, allowing continuous monitoring, analysis, planning, execution, and learning for heightened system resilience against evolving threats. Secondly, by incorporating the SDLC model, our approach ensures that security policies are embedded in every development phase, significantly reducing vulnerabilities. Our third contribution lies in providing analytical insights and measures into the model's runtime efficacy. By evaluating adaptive security metrics within the MAPE-K loop, one can demonstrate improvements in dynamically detecting, responding to, and mitigating security threats. In summary, the proposed Adaptive Software Security Model offers a systematic approach, prioritizing adaptability, proactive security measures, and continuous improvement to address the dynamic challenges of cybersecurity. It is a robust solution for organizations aiming to strengthen their software systems in the face of emerging threats in an increasingly dynamic and challenging environment.

As for future work, an exciting plan for advancing the proposed Adaptive Software Security Model involves integrating it with Continuous Integration/Continuous Delivery (CI/CD) principles in software engineering. This aim is to enhance the MAPE-K loop's dynamic nature by adapting security measures to the rapid development and deployment cycles of CI/CD pipelines with increased agility and efficient security integration. Another direction is empirical and experimental evaluations of the model in diverse software environments to determine the model's effectiveness, scalability, and adaptability in practice.


\onecolumn
\begin{table*}

\begin{tabularx}{\linewidth}{p{.15\textwidth}p{.2\textwidth}p{.65\textwidth}}
\caption{The list of metrics for evaluating the proposed adaptive software security model.} 

\label{tab:metrics}
\\
\toprule
\textbf{Metric} & \textbf{Description} & \textbf{Evaluation in MAPE-K Loop }\\
\midrule
Rate of Anomalies & Measures the frequency of detected anomalies in the system.& The monitoring phase of the MAPE-K loop observes the anomaly rate at runtime. The analyzing stage assesses patterns and trends, identifying unusual behavior. The planning phase considers adjustments to security measures based on anomaly severity. The executing step involves deploying adaptive responses to mitigate anomalies. The knowledge phase learns from previous anomalies to improve detection accuracy. \\
\midrule
Time-to-Response & Quantifies the time taken to respond to identified security threats.& The monitoring phase detects security threats. The analyzing stage assesses the urgency and severity of threats. The planning phase develops strategies for quick response. The executing step implements real-time responses. The knowledge phase captures response effectiveness for continuous improvement. \\
\midrule

Dynamic Access Controls & Evaluates the effectiveness of dynamically adjusting access controls based on contextual information.   & The monitoring phase observes user behavior and contextual factors. The analyzing phase assesses the appropriateness of access controls in various scenarios. The planning stage considers modifications to access controls based on contextual changes. The executing step implements dynamic adjustments to access controls. The knowledge phase captures the outcomes and adjusts policies for better adaptability. \\
\midrule

Incident Response Effectiveness & Measures how effectively the system responds to and contains security incidents.& The monitoring phase detects security incidents. The analyzing step assesses incident severity and impact. The planning phase formulates strategies for containment and response. The executing stage implements incident response measures. The knowledge phase captures insights from incident outcomes for continuous improvement.\\
\midrule

Adaptive Policy Success Rate & Gauges the success rate of adapting and implementing security policies in response to evolving threats. & The monitoring phase identifies emerging threats. The analyzing stage assesses the relevance of existing security policies. The planning phase devises adaptive policies based on threat analysis. The executing step implements adaptive security policies. The knowledge phase evaluates the success and effectiveness of adaptive policies over time.\\

\bottomrule
\end{tabularx}
\end{table*}
\twocolumn

\end{document}